\documentclass[twocolumn,floatfix,prx,aps,showpacs]{revtex4-1}
\usepackage{graphicx,amsmath,amssymb,color}
\usepackage{nicefrac}
\usepackage[titletoc,title]{appendix}

\newcommand{\be}{\begin{equation}}
\newcommand{\ee}{\end{equation}}

\newcommand{\ba}{\begin{eqnarray}}
\newcommand{\ea}{\end{eqnarray}}

\begin{document}

\title{Particle-hole symmetry for composite fermions: An emergent symmetry in the fractional quantum Hall effect}
\author{Ajit C. Balram$^{1}$ and J. K. Jain$^{2}$}
\affiliation{
   $^{1}$Niels Bohr International Academy and the Center for Quantum Devices,
Niels Bohr Institute, University of Copenhagen, 2100 Copenhagen, Denmark}
\affiliation{
   $^{2}$Department of Physics, 104 Davey Lab, Pennsylvania State University, University Park, Pennsylvania 16802, USA}
\date{\today}

\begin{abstract} 
The particle-hole (PH) symmetry of {\em electrons} is an exact symmetry of the electronic Hamiltonian confined to a specific Landau level, and its interplay with the formation of composite fermions has attracted much attention of late. This article investigates an emergent symmetry in the fractional quantum Hall effect, namely the PH symmetry of {\em composite fermions}, which relates states at composite fermion filling factors $\nu^*=n+\bar{\nu}$ and $\nu^*=n+1-\bar{\nu}$, where the integer $n$ is the $\Lambda$ level index and $0\leq \bar{\nu}\leq 1$. Detailed calculations using the microscopic theory of composite fermions demonstrate that for low lying $\Lambda$ levels (small $n$): (i) the 2-body interaction between composite-fermion particles is very similar, apart from a constant additive term and an overall scale factor, to that between composite-fermion holes in the same $\Lambda$ level; and (ii) the 3-body interaction for composite fermions is an order of magnitude smaller than the 2-body interaction. Taken together, these results imply an approximate PH symmetry for composite fermions in low $\Lambda$ levels, which is also supported by exact diagonalization studies and available experiments. This symmetry, which relates states at electron filling factors $\nu={n+\bar{\nu}\over 2(n+\bar{\nu})\pm 1}$ and $\nu={n+1-\bar{\nu}\over 2(n+1-\bar{\nu})\pm 1}$, is not present in the original Hamiltonian and owes its existence entirely to the formation of composite fermions.  With increasing $\Lambda$ level index, the 2-body and 3-body pseudopotentials become comparable, but at the same time they both diminish in magnitude, indicating that the interaction between composite fermions becomes weak as we approach $\nu=1/2$.
\end{abstract}
\maketitle

\section{Introduction}

The Hamiltonian of two-dimensional electrons subjected to a perpendicular magnetic field and interacting via a 2-body interaction satisfies an exact symmetry, namely the particle-hole (PH) symmetry, in the limit when the cyclotron energy is infinitely large compared to the interaction strength. The PH symmetry of electrons implies that for fully spin polarized electrons the eigenstates at $\nu$ and $1-\nu$ are related by PH transformation and the eigenspectra at $\nu$ and $1-\nu$ are identical up to an overall additive constant. How this symmetry enters into the composite fermion(CF) theory\cite{Jain07,Jain89,Lopez91,Halperin93} is a subtle issue, because composite fermions are complex emergent particles and it is unclear what symmetry must be imposed on composite fermions to guarantee the PH symmetry of the underlying {\em electrons}. Computer calculations have nonetheless shown that the Jain CF states at $\nu=n/(2n+1)$ and $\nu=1-n/(2n+1)=(n+1)/(2n+1)$ are related by PH transformation of electrons to an excellent approximation\cite{Wu93,Jain07,Balram16b}, and the CF wave function of the Halperin-Lee-Read Fermi sea \cite{Halperin93} at the half filled Landau level (LL) also has a high overlap with its hole conjugate\cite{Rezayi00}. (These results also follow from the observation that these wave functions are very close to the Coulomb eigenstates in the lowest Landau level (LLL)\cite{Dev92,Jain97,Jain07,Balram13,Jain14}.)  A new field theoretical description by Son\cite{Son15} treats  composite fermions as Dirac particles to make the PH symmetry of electrons manifest; many subsequent articles have further developed this theory\cite{Wang15b,Metlitski15,Wang15,Geraedts15,Murthy15,Liu15a,Kachru15,Metlitski15a,Potter16,Mross16,Wang16,Mulligan16,Levin16,Nguyen16,Balram16b,Nguyen17,Yang17}. Wang {\em et al.} \cite{Wang17} have shown that when properly calculated, the composite fermion Chern-Simons theory \cite{Halperin93} also produces observables consistent with PH symmetry.  

These examples show that PH symmetry of electrons remains valid in spite of the formation of composite fermions. The present article deals with a symmetry that is not present in the original Hamiltonian but arises entirely due to the formation of composite fermions, namely the PH symmetry of {\em composite fermions}. Specifically, we ask to what extent composite fermions confined to a given $\Lambda$ level ($\Lambda$L), which are analogous to electron Landau levels, satisfy the PH symmetry, i.e., how accurately are the states at the CF fillings of $\nu^*=n+\bar{\nu}$ and $\nu^*=n+1-\bar{\nu}$ related by PH symmetry of composite fermions. If present, this would be an emergent symmetry, arising entirely due to the formation of composite fermions, because no symmetry of the original electron Hamiltonian relates electron states at $\nu={n+\bar{\nu}\over 2p(n+\bar{\nu})\pm 1}$ and $\nu={n+1-\bar{\nu}\over 2p(n+1-\bar{\nu})\pm 1}$.

\section{Criteria for PH symmetry of composite fermions}

The PH symmetry of electrons can be deduced from the interaction Hamiltonian when LL mixing is absent. Such a simple criterion is not available for composite fermions. In particular, it is not possible to achieve a limit where the $\Lambda$ level separation is large compared to inter-CF interaction, as they both arise from the same Coulomb interaction between the underlying electrons. PH symmetry for composite fermions can surely not be an exact symmetry of the problem. We therefore define PH symmetry for composite fermions in an operational sense. Composite fermions would satisfy PH symmetry within a given $\Lambda$L provided the following two conditions are met:

\begin{enumerate}

 \item The 2-body interactions between two CF particles and between two CF holes are same in the $\Lambda$ level in question, modulo a constant additive term and a scale factor (which do not affect the eigenstates). There is no {\em \'a priori} reason to expect this to be the case, and only detailed quantitative calculation can address this issue.
 
 \item The inter-CF interaction for a given $\Lambda$L is dominated by the purely 2-body interaction, with 3- and higher body terms, which break PH symmetry, being negligible. 
 
\end{enumerate}

Should these conditions be satisfied, the descriptions of the state at any given filling factor $\nu^*=n+\bar{\nu}$ in terms of (i)  CF particles on top of the $\nu^*=n$ state and (ii) CF holes on top of the $\nu^*=n+1$ state would be equivalent, implying a PH symmetry for composite fermions. 

Because composite fermions are emergent collective particles with complex internal structure, the interaction between them is also complex. Nonetheless, a determination of the inter-CF interaction is possible using the microscopic formulation of composite fermions. The composite fermions inherit their interaction from the Coulomb interaction between electrons, but their interaction depends strongly on their structure, which, in turn, varies significantly with the filling factor of the background state. We determine below the 2- and 3-body interaction pseudopotentials for composite fermions in various $\Lambda$ levels to address the above two criteria. The 2-body pseudopotentials have been determined in the past \cite{Sitko96,Wojs00,Lee01,Lee02,Wojs04,Balram16c} but it would be necessary to obtain them with greater accuracy for the current purposes. 

We shall use for our calculations Haldane's spherical geometry\cite{Haldane83,Greiter11} in which $N$ electrons reside on the surface of a sphere which is radially pierced by a total magnetic flux of $2Q$ in units of the magnetic flux quantum $\phi_0=hc/e$. The effective flux seen by composite fermions is denoted by $2Q^{*}$ and is given by the relation $2Q^{*}=2Q-2p(N-1)$. Composite fermions form their own Landau-like levels, called $\Lambda$ levels, in the reduced magnetic field, and fractional quantum Hall (FQH) effect at $\nu=n/(2pn\pm 1)$ results when composite fermions form an integer quantum Hall (IQH) state with $n$ filled $\Lambda$ levels. The Jain wave function for these FQH ground states are given by\cite{Jain89,Jain07}: $\psi_{n/(2pn\pm 1)}^{\rm CF}=\prod_{i<j}(u_{i}v_{j}-u_{j}v_{i})^{2p}\Phi_{\pm n}$, where $\Phi_{n}$ ($\Phi_{-n}=[\Phi_{n}]^{*}$) is the the Slater determinant of the IQH state of $n$ filled LLs of electrons. Here $u_{i}=\cos(\theta_{i}/2)e^{i\phi_{i}/2}$ and $v_{i}=\sin(\theta_{i}/2)e^{-i\phi_{i}/2}$ are the spinor coordinates and $\theta$ and $\phi$ are the polar and azimuthal angles on the sphere respectively. The excited states can be similarly obtained from the corresponding IQH state. In this work we will only be interested in the case with $p=1$. 

\subsubsection{2-body pseudopotentials}

The 2-body pseudopotentials of composite fermions are the energies of two composite fermions in a given $\Lambda$ level in a state of definite relative angular momentum $m$. For our finite systems, we define them as\cite{Lee01,Lee02}:
\begin{equation}
V_{m}^{\rm CF(2)}=E'_{m}-E'_{m_{\rm max}}+E_{m_{\rm max}}^{\rm disc}
\label{eq_CF_two_body_pps}
\end{equation}
Here $E'_{m}=\sqrt{2Q\nu/N}E_{m}$ is the density corrected Coulomb energy\cite{Morf86} of the state with two composite fermions in relative angular momentum $m$. $E_m$ is defined as : 
\begin{equation}
E_{m}=\left(\frac{\langle \psi_{m}^{\rm CF}|\sum_{i<j}\frac{1}{r_{ij}}|\psi_{m}^{\rm CF} \rangle}{\langle\psi_{m}^{\rm CF}|\psi_{m}^{\rm CF}\rangle}-\frac{N^2}{2\sqrt{Q}} \right)~ \frac{e^{2}}{\epsilon\ell}
\end{equation}
where $|\psi_{m}^{\rm CF}\rangle$ is the wave function of composite fermions in the relative angular momentum $m$ state, and the term $-N^2/(2\sqrt{Q})~e^{2}/\epsilon\ell$ is the contribution due to the presence of the positive background charge (note that the pseudopotentials are defined in such a way that the contribution of the background positive charge drops out). The relative angular momentum of two composite fermions residing in the $\Lambda$L indexed by $n$ is given by $m=2(Q^{*}+n)-L$, where $L$ is their total orbital angular momentum. The two composite fermions are maximally separated in the state with relative angular momentum $m_{\rm max}$. The last two terms in Eq.~\ref{eq_CF_two_body_pps} are an overall additive constant chosen so that the interaction pseudopotential at $m_{\rm max}$ equals $E^{\rm disc}_{m_{\rm max}}$, given by\cite{Lee01,Lee02}:
\begin{equation}
E_{m_{\rm max}}^{\rm disc}=\frac{1}{(2n+1)^{\frac{5}{2}}}
\frac{\Gamma(m_{\rm max}+\frac{1}{2})}{2\Gamma(m_{\rm max}+1)}~ \frac{e^{2}}{\epsilon\ell}
\label{disc_Coulomb_pps}
\end{equation}
which is the energy of two particles of charge $1/(2n+1)$ in a relative angular momentum $m_{\rm max}$ state, confined to the lowest LL in an effective magnetic field. (Another possibility would be to choose $E_{m_{\rm max}}^{\rm disc}$ to be the corresponding value on the sphere, but since the spherical value is system size dependent, it is more convenient to choose the disc value. In any case, we will be considering sufficiently large systems that the difference between spherical and disc values for $E_{m_{\rm max}}^{\rm disc}$ is negligible.) With the above definitions we have $V^{\rm CF(2)}(m_{\rm max})=E_{m_{\rm max}}^{\rm disc}$ as should be the case for a large system. \\

\subsubsection{Irreducible 3-body pseudopotentials}

We are interested in calculating the ``irreducible" 3-body CF pseudopotentials denoted by $V_{m,\alpha}^{\rm CF(3),irr}$, where $m$ is the relative angular momentum of the three composite fermions\cite{Simon07} and $\alpha$ is used to index the different states with the same $m$. To obtain these we first calculate $V_{m,\alpha}^{\rm CF(3)}$, the energy of three composite fermions in the state with relative angular momentum $m$ and index $\alpha$, and subtract the contribution arising from purely 2-body interactions. \\

We calculate the $V_{m,\alpha}^{\rm CF(3)}$ using the method of composite fermion diagonalization\cite{Mandal02,Jain07}. We consider $n$ filled $\Lambda$Ls of composite fermions (indexed from $0$ to $n-1$) and three composite fermions in the $n$th $\Lambda$L. The relative angular momentum of these three composite fermions is given by $m=3(Q^{*}+n)-L$, where $L$ is their total orbital angular momentum. We define the energy of a state as:
\begin{equation}
E_{\rm 3b}(m,\alpha)=\left(\frac{\langle\psi^{3,\rm CF}_{m,\alpha}|\sum_{i<j}\frac{1}{r_{ij}}|\psi^{3,\rm CF}_{m,\alpha}\rangle}{\langle\psi^{3,\rm CF}_{m,\alpha}|\psi^{3,\rm CF}_{m,\alpha}\rangle}-\frac{N^2}{2\sqrt{Q}} \right)~ \frac{e^{2}}{\epsilon\ell}
\end{equation}
where $|\psi^{3,\rm CF}_{m,\alpha}\rangle$ is the microscopic wave function in which three composite fermions are in the relative angular momentum $(m,\alpha)$ state. Note that the Coulomb interaction between electrons is purely 2-body; the multi-body interaction between composite fermions arises due to complex many body correlations. \\

$V_{m,\alpha}^{\rm CF(3)}$ also contains contribution from 2-body interaction, which must be subtracted. We calculate the purely 2-body contribution by taking three particles at the corresponding flux of $2(Q^{*}+n)$ in the LLL and exactly diagonalizing the 2-body composite fermion pseudopotentials $V_{m}^{\rm CF(2)}$. Using these pseudopotentials we define:
\begin{equation}
E_{\rm 2b}(m,\alpha)=\sum_{1<=i<j<=3} \sum_{m_{2}} V_{m_{2}}^{\rm CF(2)} \mathcal{P}_{m_{2}}^{ij}|\psi^{3,\rm e,LLL}_{m,\alpha} \rangle
\end{equation}
where $|\psi^{3,\rm e,LLL}_{m,\alpha}\rangle$ is the wave function in which three electrons are in the relative angular momentum $(m,\alpha)$ state in the lowest Landau level and $\mathcal{P}_{m_{2}}^{ij}$ projects particles $i$ and $j$ onto a state of relative angular momentum $m_{2}$. \\

From the above mentioned quantities we define the 3-body \emph{irreducible} composite fermion pseudopotentials as:
\begin{eqnarray}
V_{m,\alpha}^{\rm CF(3),irr}&=&\left(E'_{\rm 3b}(m,\alpha)-E_{\rm 2b}(m,\alpha) \right) \nonumber \\
  &&-\left(E'_{\rm 3b}(m_{\rm max})-E_{\rm 2b}(m_{\rm max}) \right)
\end{eqnarray}
where in the term corresponding to $m_{\rm max}$ is a constant that does not affect results in any way and $E'$ as before is the density corrected energy: $E'=\sqrt{2Q\nu/N}E$. Using this definition we have $V^{\rm CF(3),irr}(m_{\rm max})=0$ for a large angular momentum $m_{\rm max}$, which is reasonable since at large values of $m$ there is no overlap between the three particles and the energy of such a state should just be given by the pair-wise Coulomb interaction of the three particles. For technical reasons, instead of setting $V^{\rm CF(3),irr}(m_{\rm max})=0$, we choose to set $V^{\rm CF(3),irr}(m=10)=0$. (For larger $m$, the statistical error from Monte Carlo calculation is large, as the number of Slater determinants participating in the wave function increases with the relative angular momentum $m$.)

We have described the 2- and 3-body pseudopotentials for CF particles. It is straightforward to generalize it to CF holes and also to include the spin degree of freedom. \\

\section{Results}
\label{sec:results}

\begin{figure}
\begin{center}
\includegraphics[width=7.5cm,height=5.0cm]{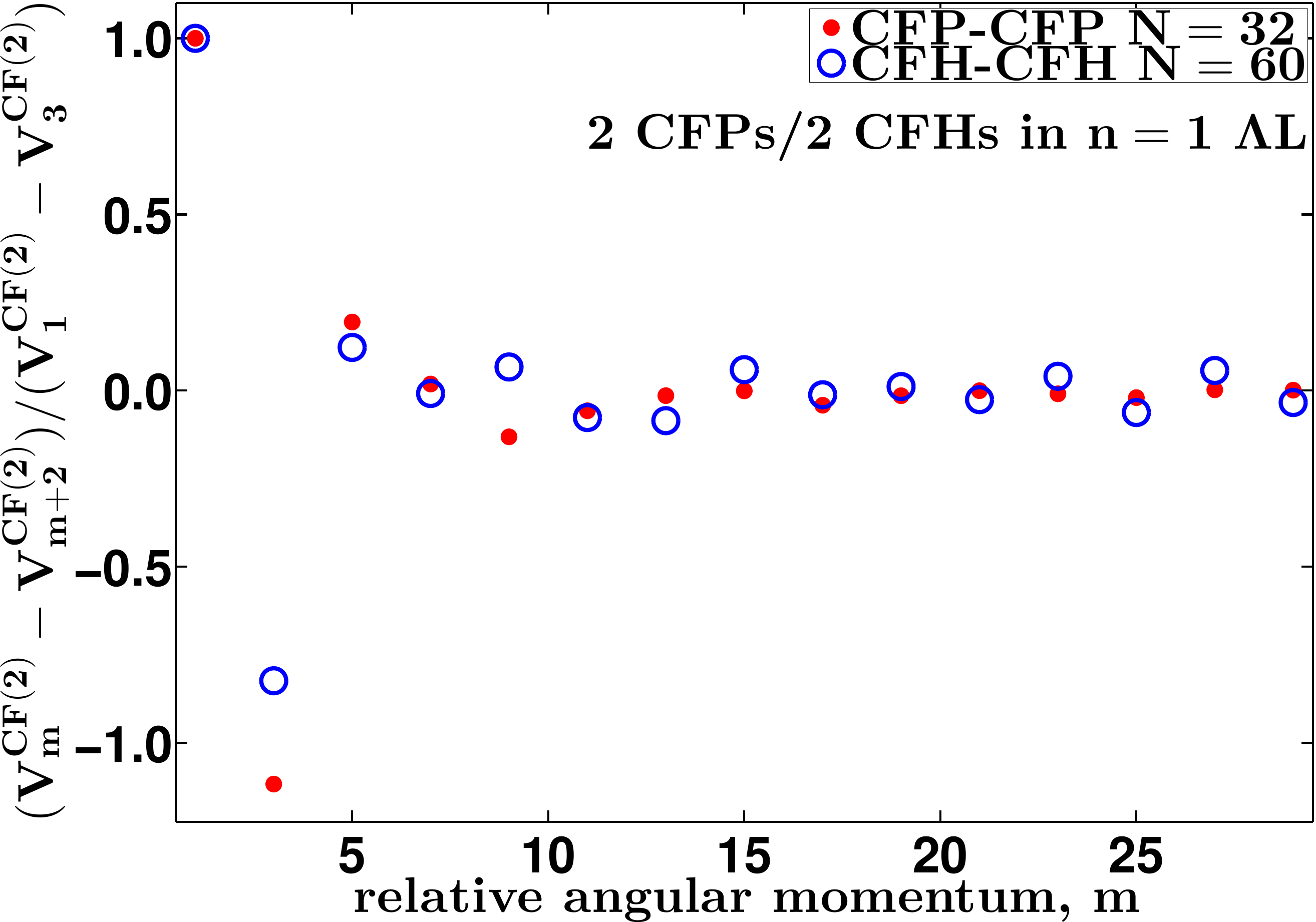} 
\includegraphics[width=7.5cm,height=5.0cm]{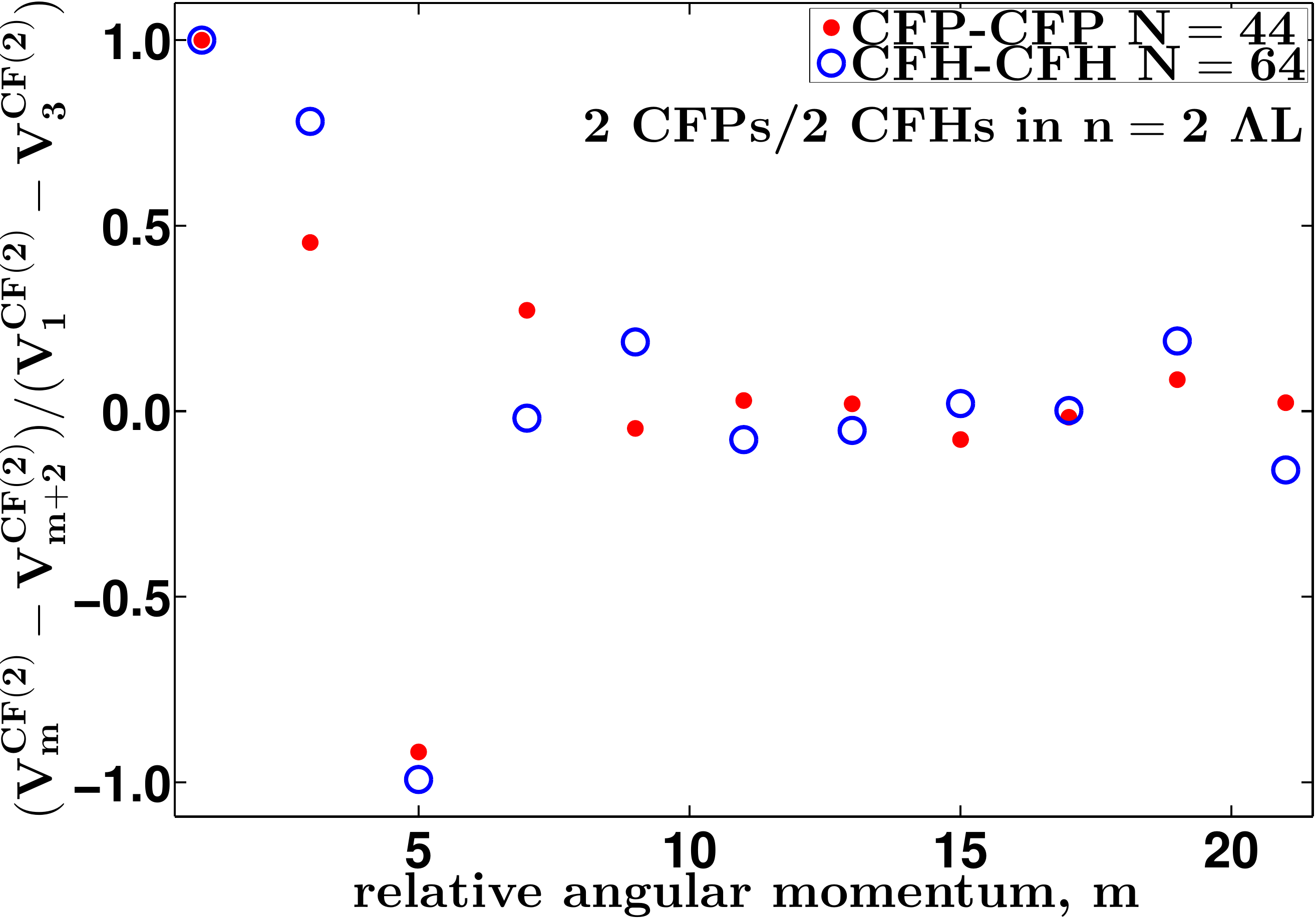} 
\includegraphics[width=7.5cm,height=5.0cm]{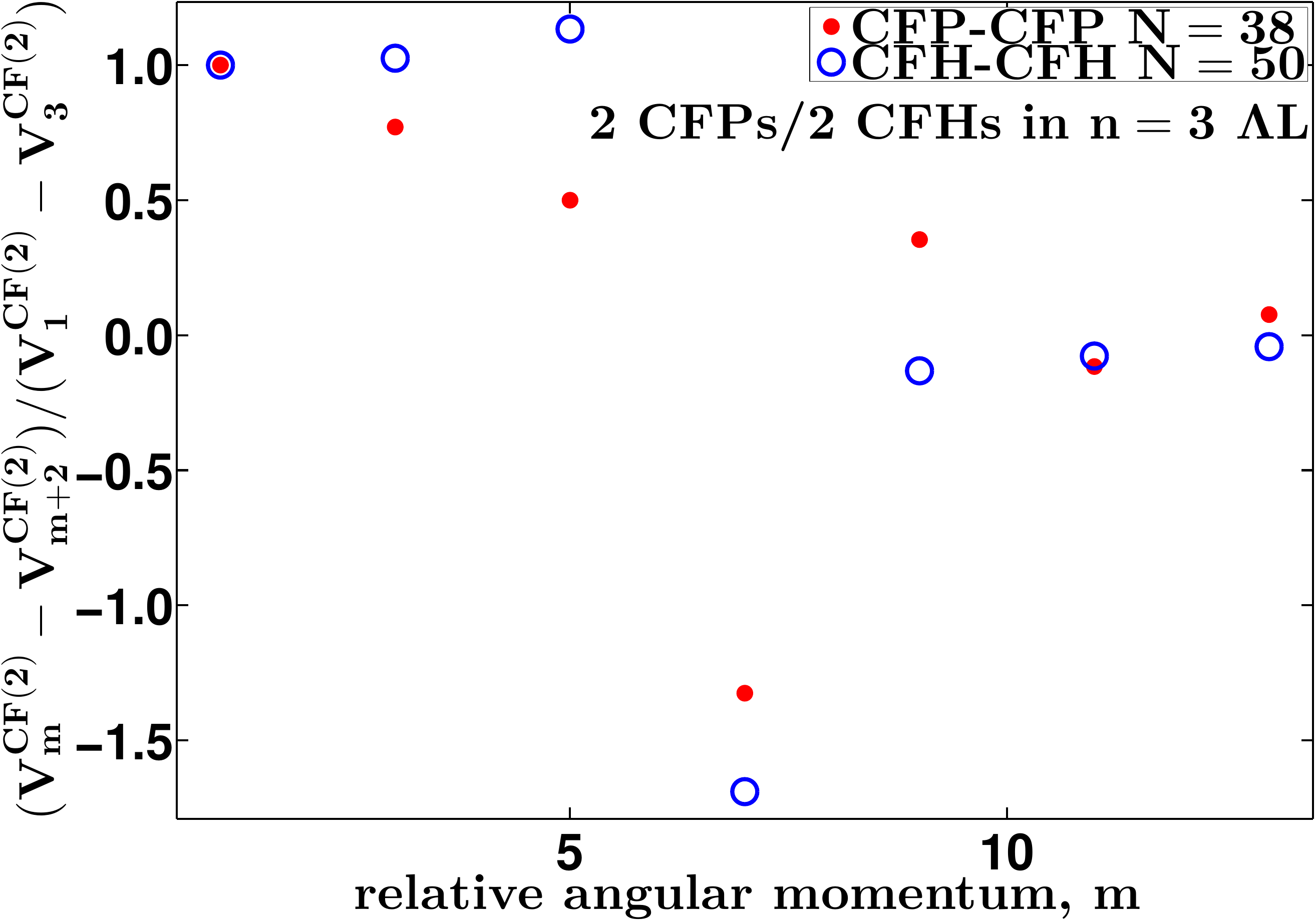} 
\includegraphics[width=7.5cm,height=5.0cm]{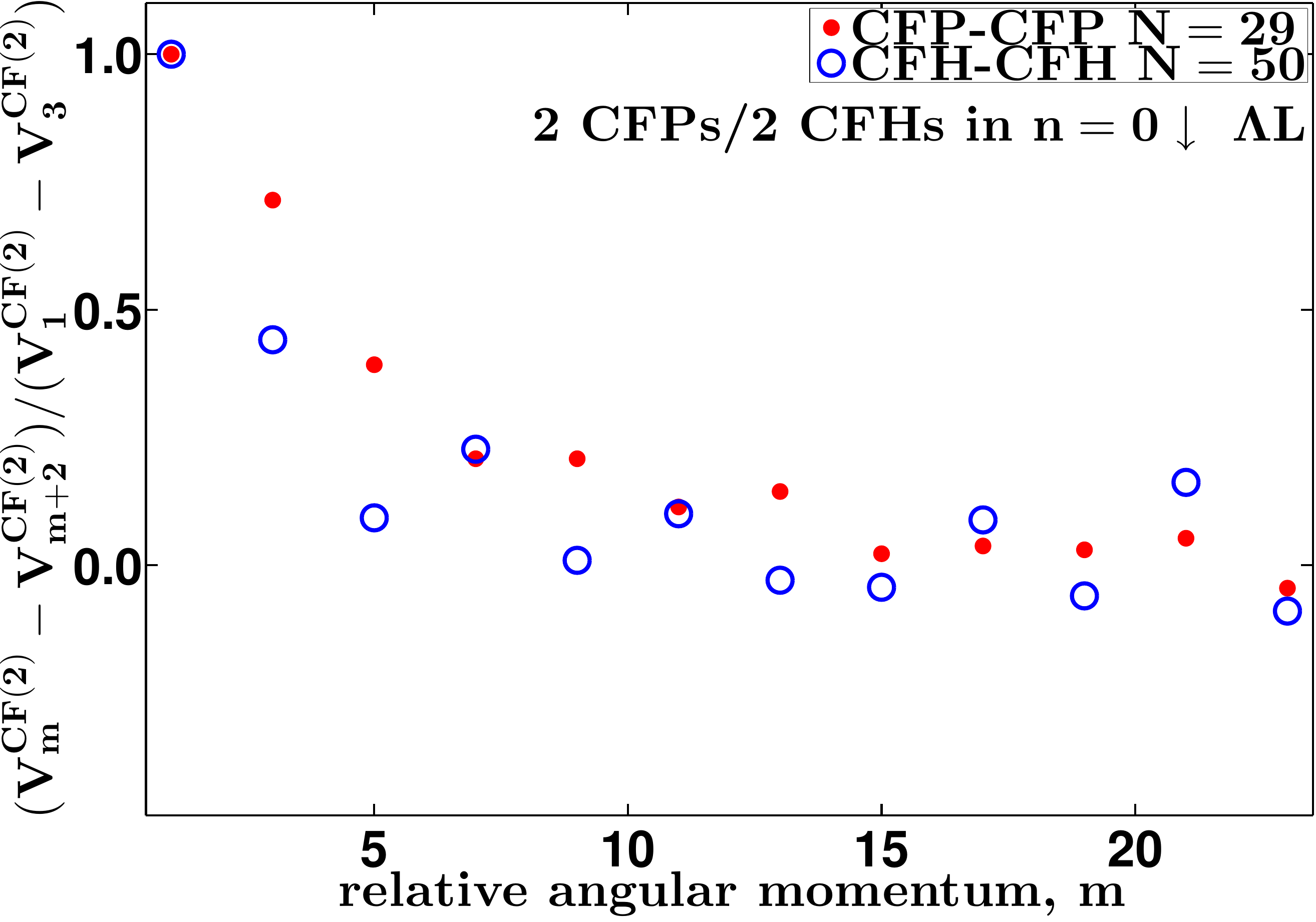}
\caption{(color online) Comparison of the 2-body CF particle (CFP) and CF hole (CFH) pseudopotentials for fully spin polarized composite fermions in the $n=1$ (top panel), $n=2$ (second panel), $n=3$ (third panel) and $n=0\downarrow$ (bottom panel) $\Lambda$L. The bottom right panel shows results for spinful composite fermions in the $n=0\downarrow$ $\Lambda$L with the $n=0\uparrow$ $\Lambda$L fully occupied. The pseudopotentials are obtained from the microscopic wave function of composite fermions in the spherical geometry as explained in the main text. The quantity $(V_{m}^{\rm CF(2)}-V_{m+2}^{\rm CF(2)})/(V_{1}^{\rm CF(2)}-V_{3}^{\rm CF(2)})$ is plotted because it is invariant under an overall additive constant and a change of the overall scale.}
\label{fig:two_body_CFP_CFH_pps}
\end{center}
\end{figure}

We first compare the 2-body pseudopotentials for CF particles and CF holes.  In Appendix \ref{app:two_body_CF_pps} we have tabulated the 2-body pseudopotentials of CF particles and CF holes residing in different $\Lambda$Ls indexed by $n$. The 2-body pseudopotentials of fully polarized composite fermions have a maximum at relative angular momentum $m=2n+1$. Because the actual eigenstates are not affected by an overall additive term or by a scale factor, we compare the two pseudopotentials by plotting the quantity: $(V_{m}^{\rm CF(2)}-V_{m+2}^{\rm CF(2)})/(V_{1}^{\rm CF(2)}-V_{3}^{\rm CF(2)})$ in Fig.~\ref{fig:two_body_CFP_CFH_pps}. These ratios for CF particles and CF holes are qualitatively very similar, and even quantitatively very close for low values of the $\Lambda$L index. Deviation between them grows with increasing $\Lambda$L index.

\begin{figure}
\begin{center}
\includegraphics[width=7.5cm,height=5.0cm]{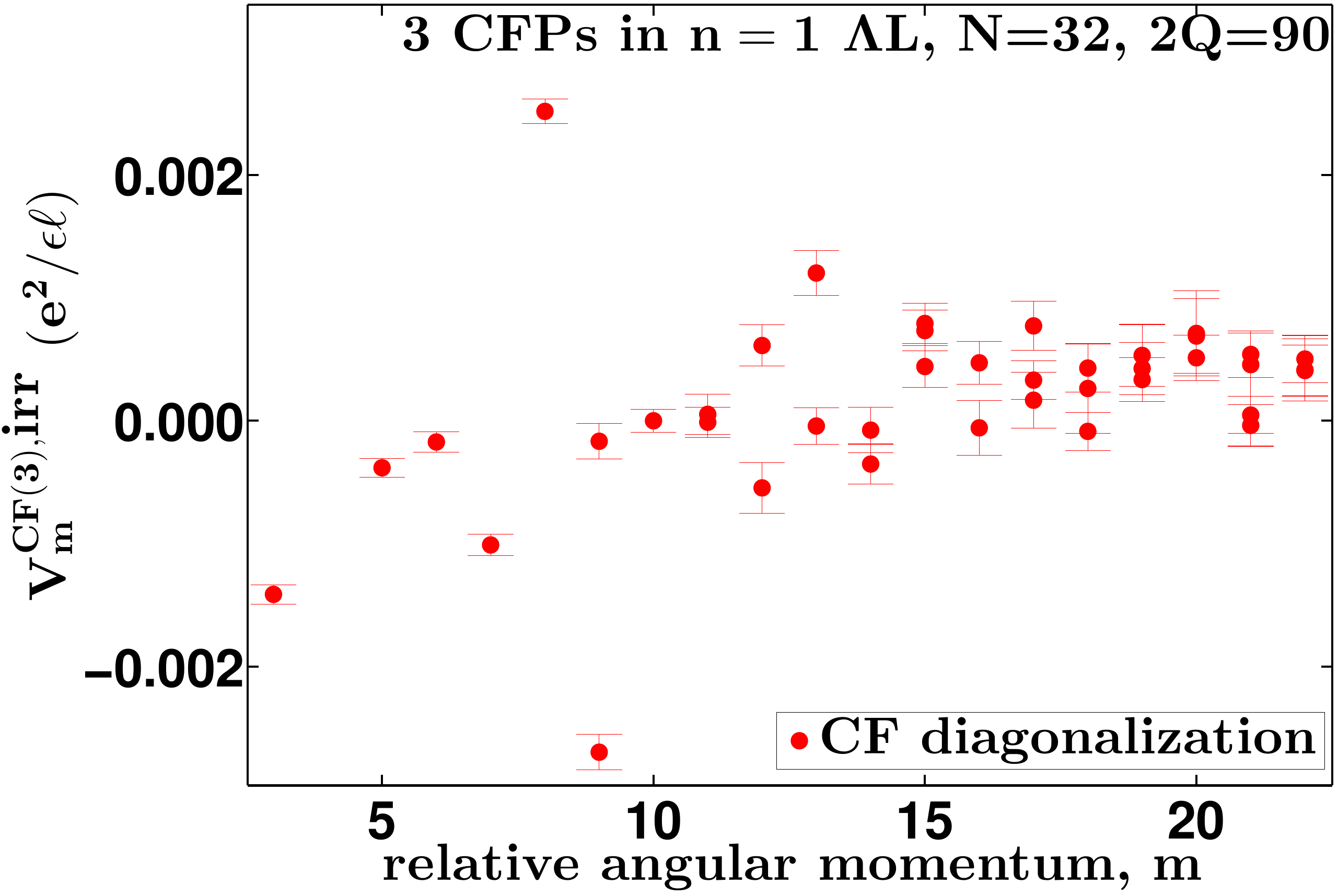} 
\includegraphics[width=7.5cm,height=5.0cm]{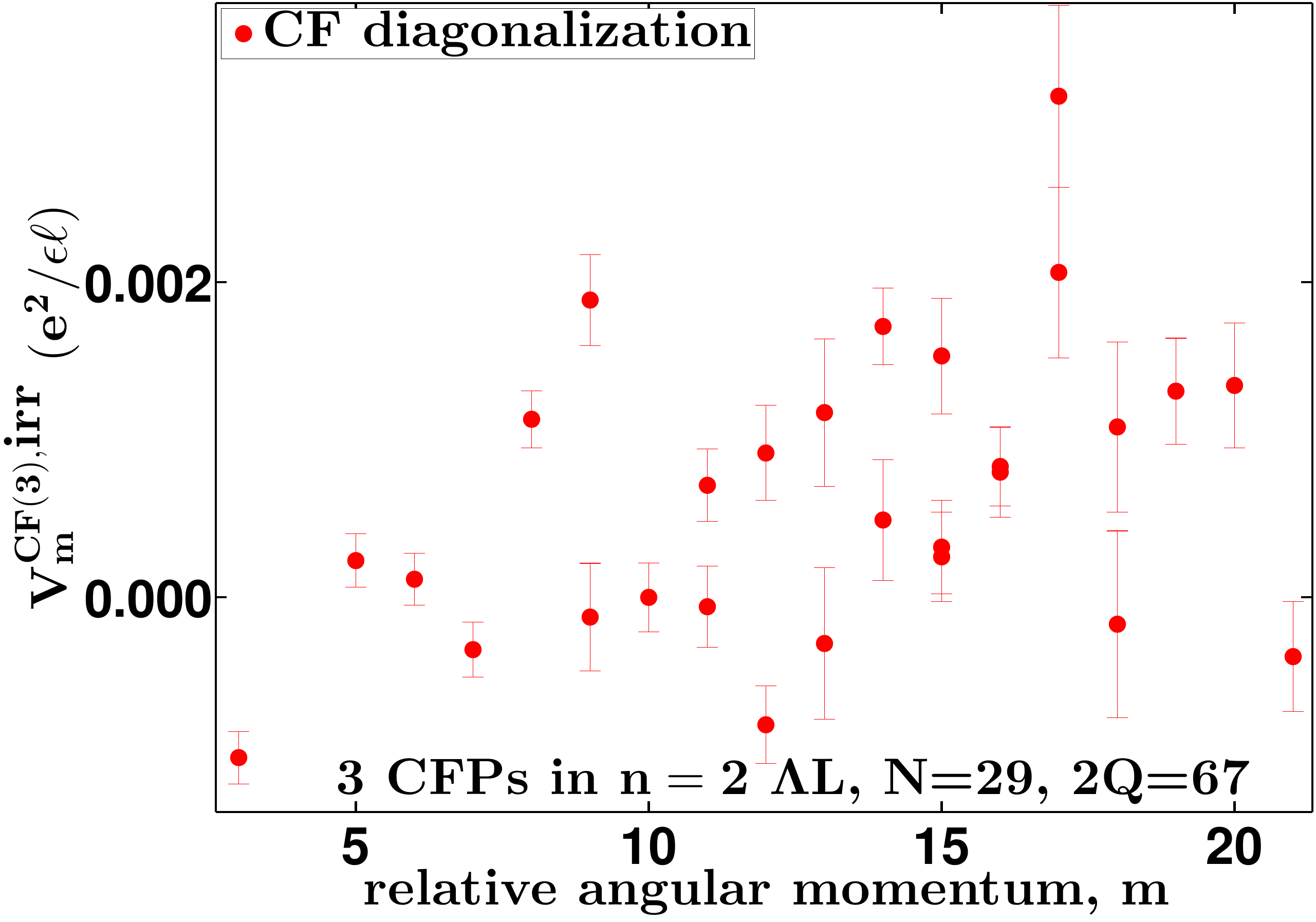}
\includegraphics[width=7.5cm,height=5.0cm]{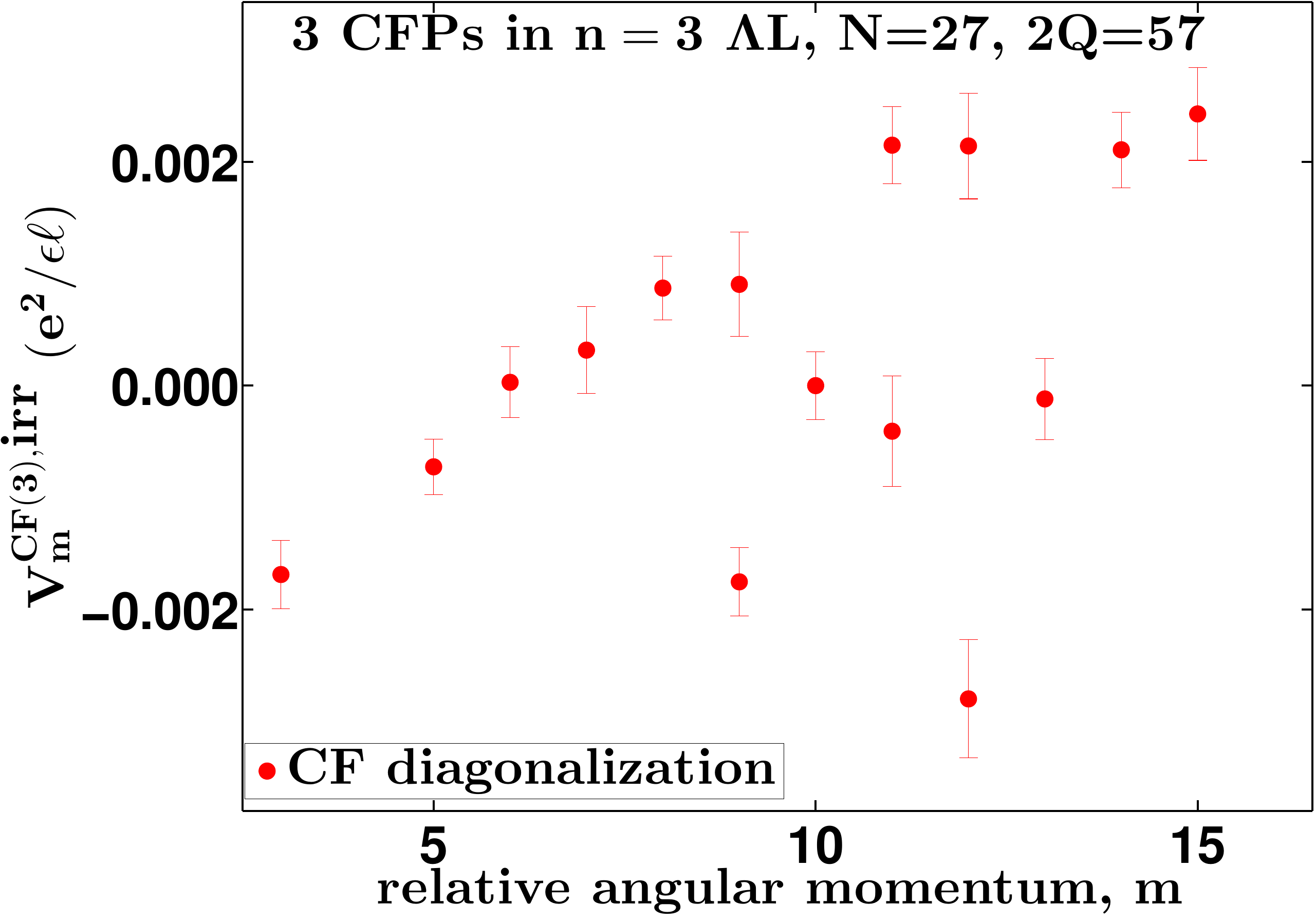} 
\includegraphics[width=7.5cm,height=5.0cm]{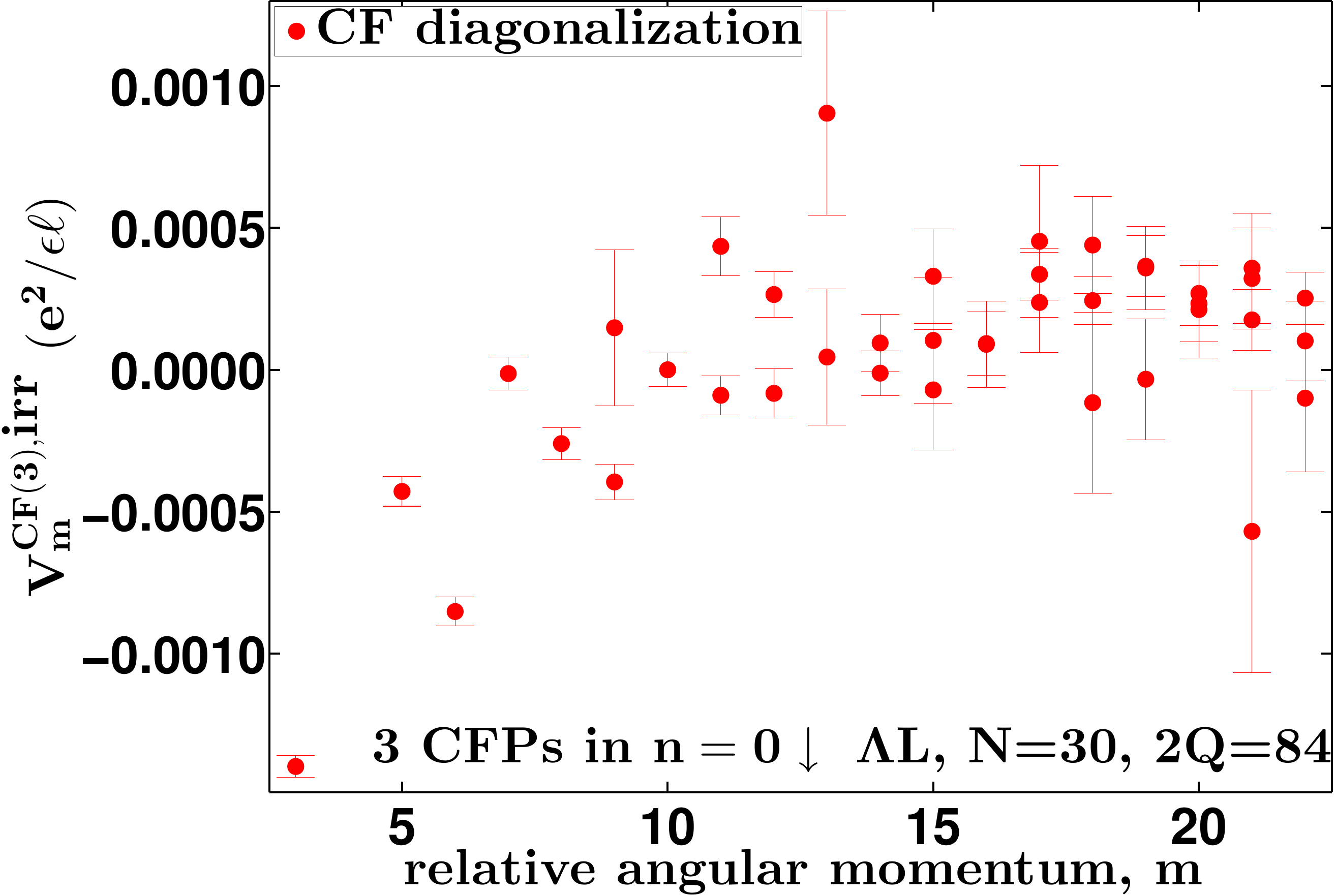}  
\caption{(color online) 3-body irreducible pseudopotentials CF particle (CFP) in the $n=1$ (top panel), $n=2$ (second panel), $n=3$ (third panel) and $n=0\downarrow$ (bottom panel) $\Lambda$L. These are defined in the main text, and are obtained from the microscopic wave function of composite fermions in the spherical geometry. In these pseudopotentials we have set $V^{\rm CF(3),irr}_{10}=0$ to fix the constant part and shown values only up to $m=22$. The error bars reflect statistical uncertainty the Monte Carlo sampling. }
\label{fig:three_body_irr_CF_pps}
\end{center}
\end{figure}

In Fig.~\ref{fig:three_body_irr_CF_pps} we show the irreducible part of the 3-body pseudopotentials for three CF particles residing in a given $\Lambda$L. Results for CF holes are similar. The smallest pseudopotential is attractive as can be anticipated from general screening arguments. Comparison with the 2-body pseudopotentials given in Table~\ref{tab:two_body_CFP_pps} shows that the irreducible 3-body pseudopotentials are an order of magnitude smaller than their 2-body counterparts for $n=1$ $\Lambda$L, and thus are quantitatively negligible. As a result, we conclude that PH symmetry of composite fermions is an approximately valid symmetry in the $n=1$ $\Lambda$ level. With increasing $\Lambda$L index $n$, the 3-body and 2-body pseudopotentials become comparable in strength, but at the same time both grow weaker compared to the 2-body pseudopotentials in the $n=1$ $\Lambda$L. The suppression of inter-CF interaction with increasing $\Lambda$L index is not surprising given that (i) the size of a CF quasiparticle or a CF quasihole grows with $n$, and (ii) the local charge decreases with $n$ (given by $1/(2n+1)$ of an electron charge). This implies that the composite fermions become more weakly interacting as we go to higher $n$, i.e. closer to $\nu=1/2$.  This is consistent with the observation \cite{Balram13} that more and more well defined bands appear in exact spectra as we go to states at $n/(2n+1)$ with higher $n$.

\begin{figure*}
\begin{center}
\includegraphics[width=7.5cm,height=5.5cm]{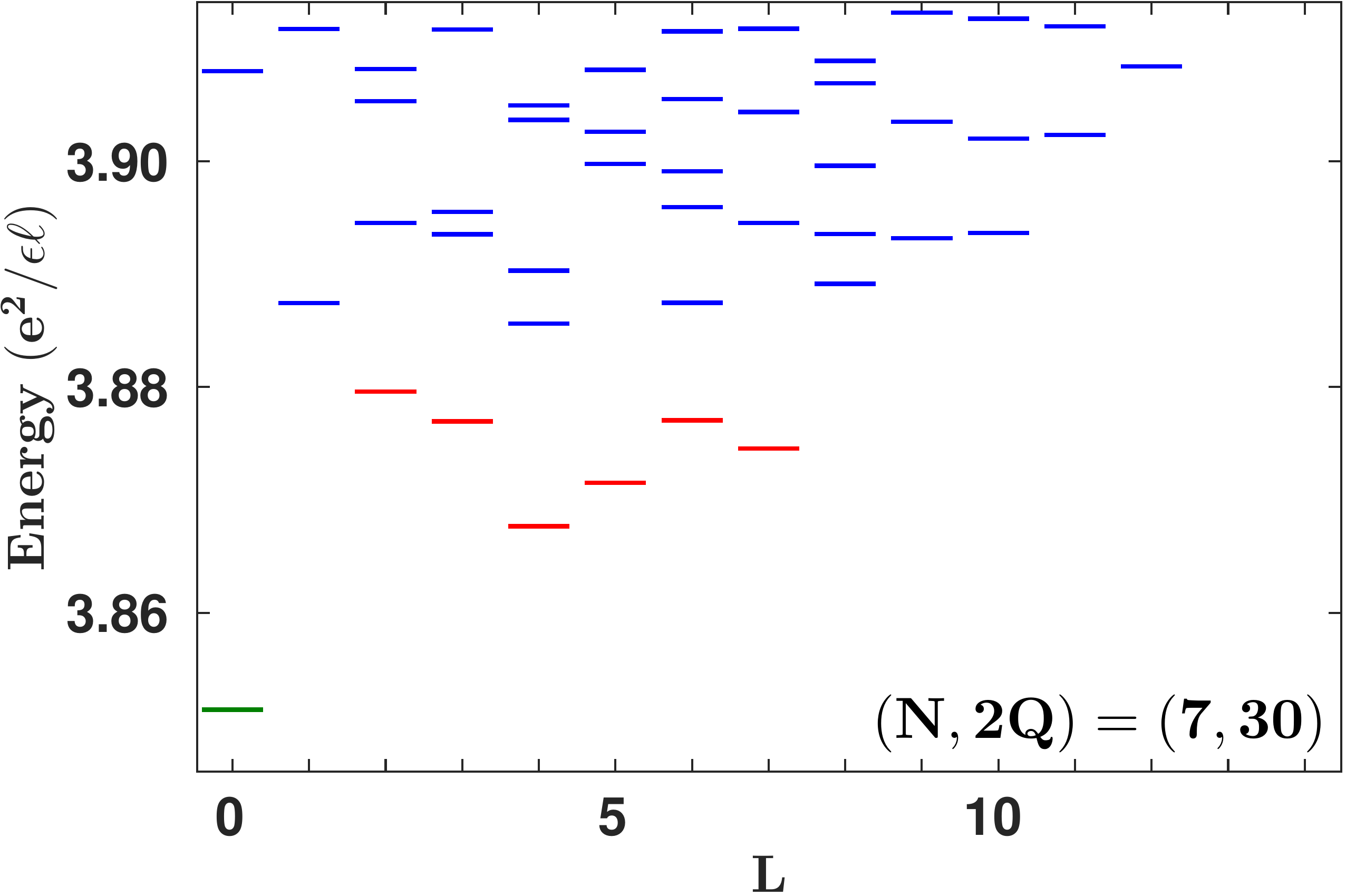} \includegraphics[width=7.5cm,height=5.5cm]{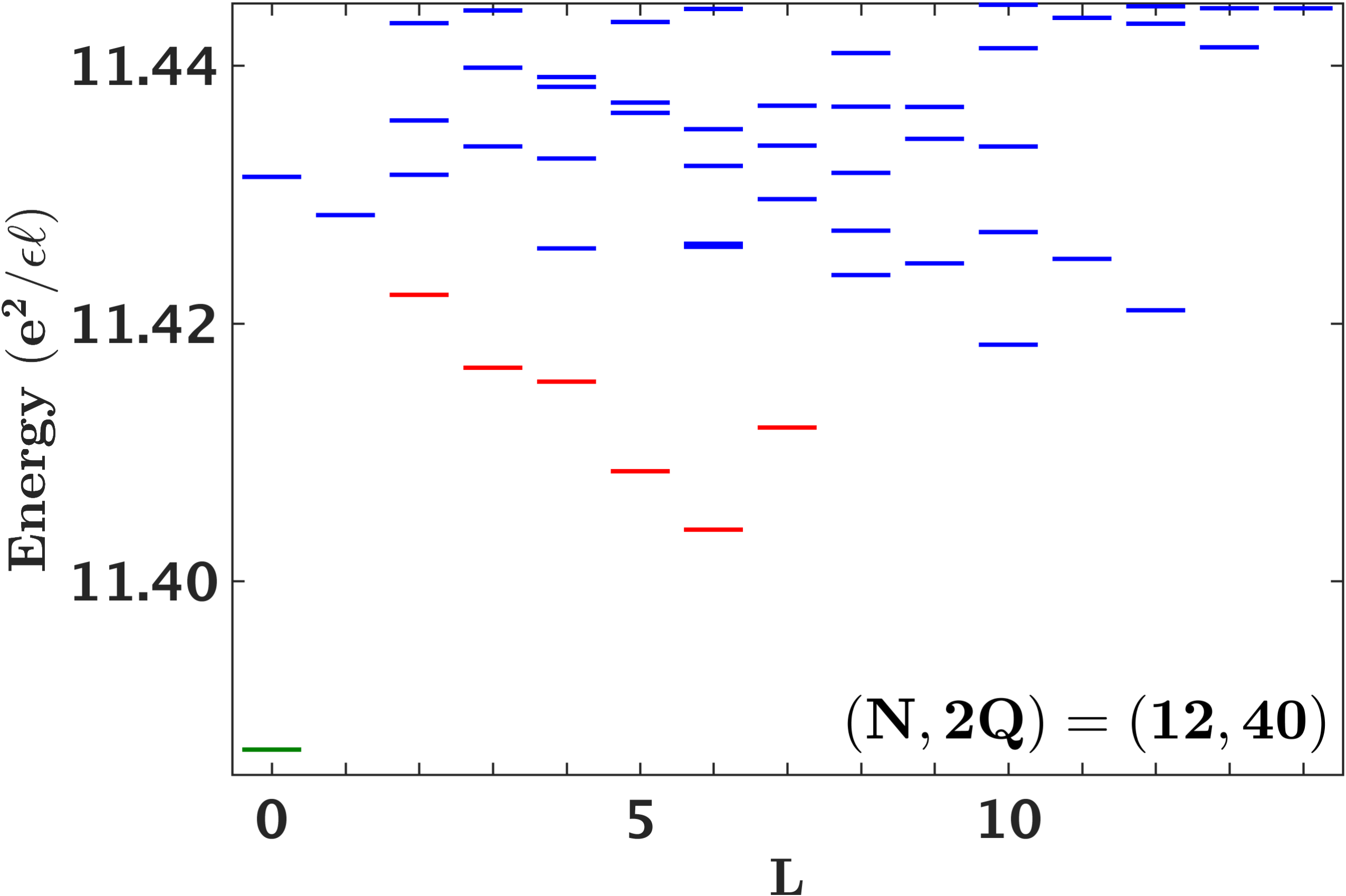} 
\includegraphics[width=7.5cm,height=5.5cm]{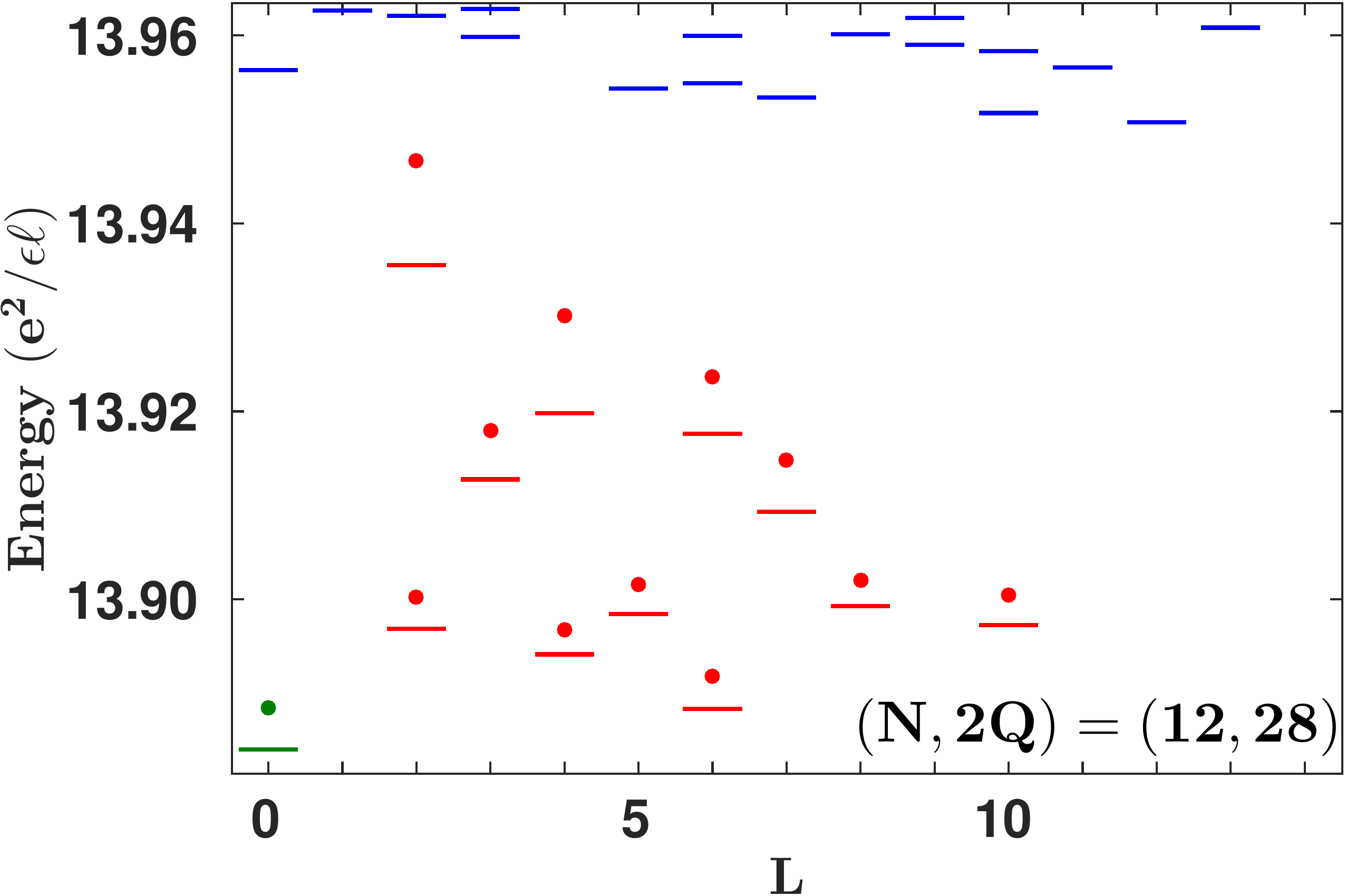} \includegraphics[width=7.5cm,height=5.5cm]{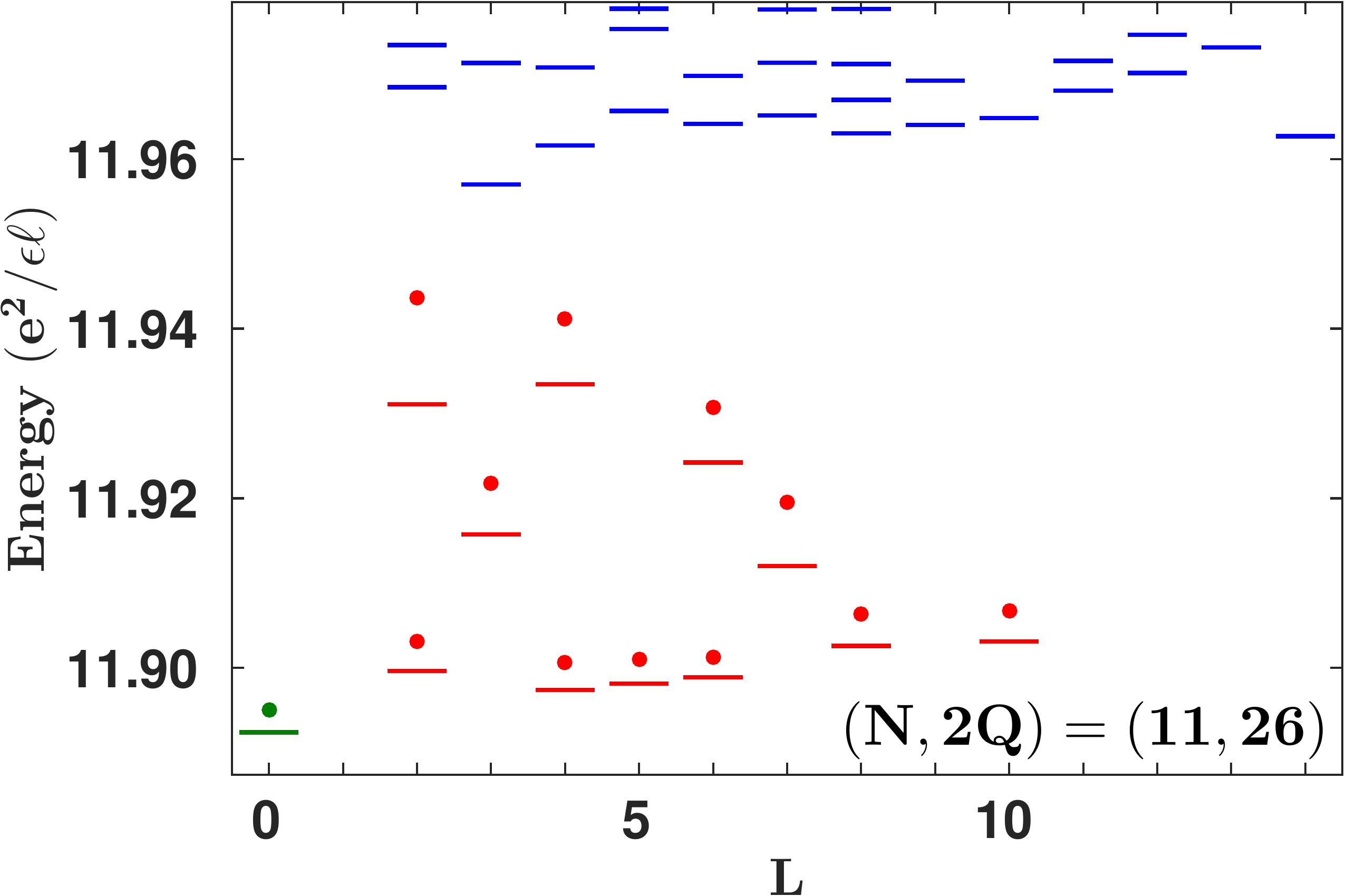} 
\includegraphics[width=7.5cm,height=5.5cm]{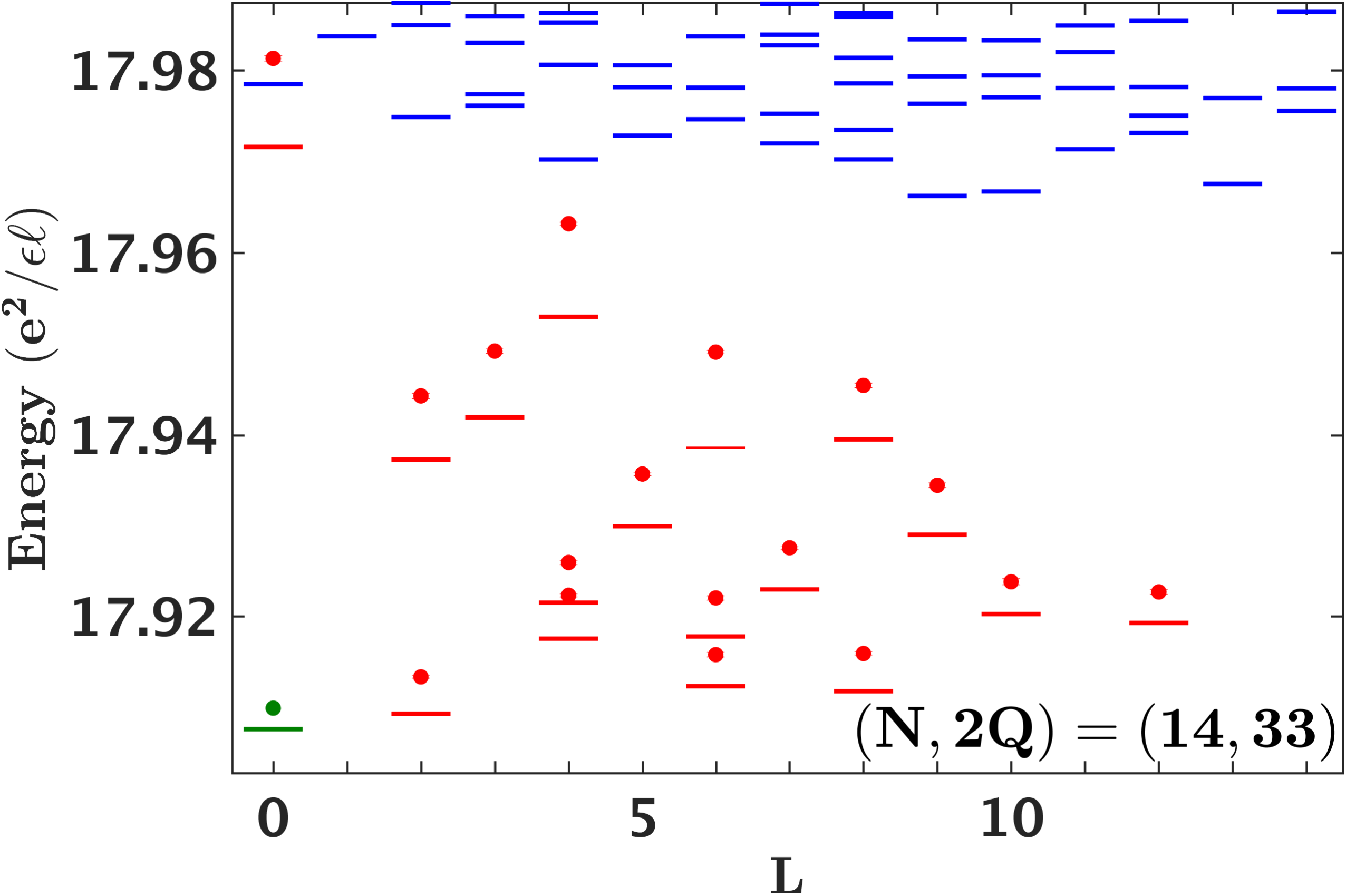} \includegraphics[width=7.5cm,height=5.5cm]{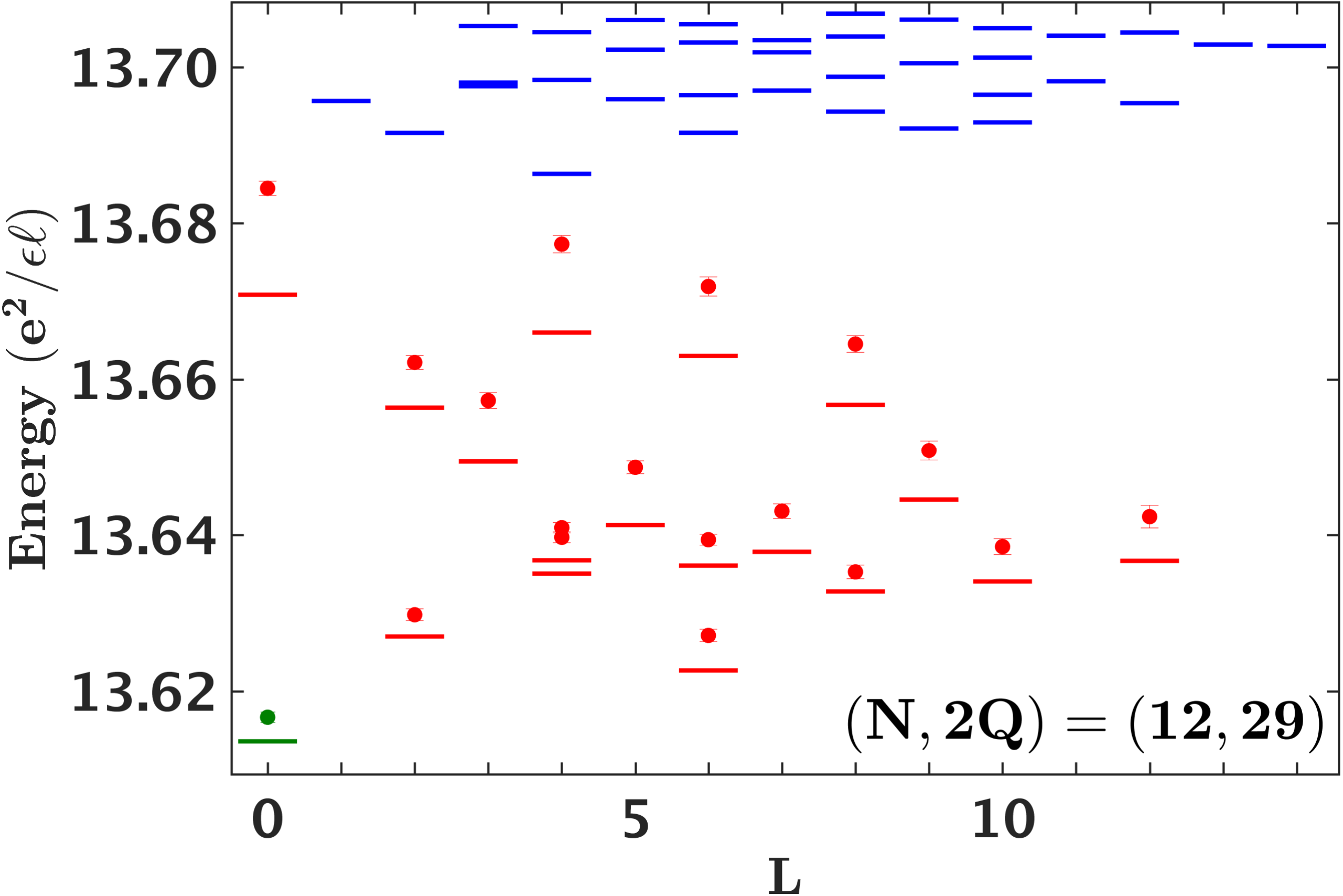} 
\caption{(color online) Coulomb spectra obtained in the spherical geometry using exact (dashes) and composite fermion (dots) diagonalization. $N$ is the number of electrons, $2Q$ is the number of flux quanta ($\phi_0=hc/e$) they are exposed to, and $L$ is their total orbital angular momentum. The right panels are related to the left panels by PH symmetry of composite fermions. To identify the various bands, the ground state is shown in green, the band of first excited states in red and rest of the energies in blue. (a) The top panels show systems with $(N,2Q)=(7,30)$ and $(12,40)$. These correspond to CF systems with $(N,2Q^*)=(7,18)$ and $(12,18)$, which have, respectively, 7 and 12 electrons in the lowest $\Lambda$L and are thus related by PH symmetry of composite fermions in the lowest $\Lambda$L which has a degeneracy of 19. (b) The middle panels show systems with $(N,2Q)=(12,28)$ and $(11,26)$ (reproduced from Mukherjee {\em et al.}\cite{Mukherjee14}). These correspond to CF systems with $(N,2Q^*)=(12,6)$ and $(11,6)$, which have, respectively, 5 and 4 electrons in the second $\Lambda$L and are thus related by PH symmetry of composite fermions in the second $\Lambda$L which has a degeneracy of 9. (c) Similarly, the two bottom panels correspond to CF systems $(N,2Q^*)=(14,7)$ and $(12,7)$, with 6 and 4 composite fermions in the second $\Lambda$L, which has a degeneracy of 10 (reproduced from Mukherjee {\em et al.}\cite{Mukherjee12}).}
\label{fig:spectra_p_h_symmetry_preserving}
\end{center}
\end{figure*}

PH symmetry of composite fermions can be tested directly in exact computer experiments by comparing exact Coulomb spectra of interacting electrons for parameters related by PH symmetry of composite fermions. Fig.~\ref{fig:spectra_p_h_symmetry_preserving} shows several such spectra, with the spectra related by PH symmetry of composite fermions shown side by side on the same row. The following aspects are noteworthy: (i) There are clearly identifiable low energy bands in each spectrum. The low energy bands of the paired spectra have a one to one correspondence, in that they have the same number of eigenstates at each $L$. (ii) The ground state occurs at the same $L$ quantum number for the paired spectra. (iii) The splitting of other states is not identical in the paired spectra, implying that the PH symmetry of composite fermions is not exact. We have seen similar behavior for many other spectra of interacting electrons which are related by PH symmetry of composite fermions. These results agree with our above conclusion that the PH symmetry of composite fermions indeed is an emergent symmetry in FQH effect, although it is not an exact symmetry.

There is actually another way of defining PH symmetry for composite fermions. We construct states of non-interacting electrons at the effective flux $2Q^*$, which, for the paired systems on the same row of Fig.~\ref{fig:spectra_p_h_symmetry_preserving}, are related by PH symmetry.  We next composite fermionize these states in the standard manner, and calculate their Coulomb energies \cite{Jain89,Jain07,Jain97,Jain97b}. (If there is more than one basis function at a given $L$, then we diagonalize the Coulomb interaction within that subspace\cite{Mandal02}.) The resulting spectra are shown by blue dots. The excellent agreement illustrates a ``hidden" PH symmetry that is much more accurate than what is reflected in the energy spectra and even captures the details and differences in how the states of the lowest band split.

An interesting example of the breaking of the PH symmetry of composite fermions possibly occurs for the fully and partially spin polarized states at $\nu=3/8$, which map into $\nu^*=1+1/2$ of fully and partially polarized composite fermions. It was proposed in Refs.~\cite{Mukherjee12} and \cite{Mukherjee14c} based on CF diagonalization studies that the $1/2$ state is an anti-Pfaffian rather than Pfaffian. We have found that our 2- and 3-body interactions in the $n=1$ $\Lambda$L and $n=0$ $\downarrow\Lambda$L determined above cannot discriminate between the Pfaffian and the anti-Pfaffian states within numerical uncertainty.  This is not surprising given that even a 3-body interaction is not always very effective in breaking PH symmetry\cite{Sreejith17} and the competition between the Pfaffian and the anti-Pfaffian can be rather subtle\cite{Rezayi17}. 

A remark regarding connection to experiments is in order. The structure of states in the filling factor range $1 < \nu^* < 2$ has been investigated in Ref.~\cite{Pan03,Pan15,Samkharadze15b}. They find evidence for FQHE at both 4/11 and 5/13, which are related by PH symmetry of composite fermions in the second $\Lambda$L, mapping into $\nu^*=1+1/3$ and $\nu^*=2-1/3$, respectively. Similarly, they find evidence for both 4/13 and 5/17 which are also related by PH symmetry of composite fermions. However, it is also clear that the PH symmetry is not exact: the 4/11 appears to be stronger than 5/13; and there is evidence for a weak 6/17 but none for 9/23.

\section{Conclusion}

We have highlighted in this article an emergent symmetry that approximately relates states at $\nu={n+\bar{\nu}\over 2(n+\bar{\nu})\pm 1}$ and $\nu={n+1-\bar{\nu}\over 2(n+1-\bar{\nu})\pm 1}$. This symmetry is not present in the original Hamiltonian and arises entirely due to the formation of composite fermions. It reflects a PH symmetry of composite fermions, which, in turn, follows from facts that for small $n$: (i) the 2-body interactions between the CF electrons and CF holes are approximately equivalent; and (ii) the 2-body interaction dominates 3- and higher body interactions.

We close the article by mentioning other examples of symmetries that owe their existence to the formation of composite fermions. Geraedts {\em et al.} \cite{Geraedts17} find that for spinful bosons in the LLL, the Jain states at $\nu$ and $2-\nu$ are related approximately by PH transformation, even though no such symmetry is present in the bosonic Hamiltonian.  Another emergent symmetry implies a one-to-one correspondence between the physics at filling factors given by $\nu=\nu^*/(2p\nu^*\pm 1)$ for different values of $p$ and different signs $\pm$, which are all related through their mapping into filling factor $\nu^*$ of composite fermions. This emergent symmetry manifests in experiments, for example, through appearance of FQH effect at the sequences $\nu=n/(4n\pm 1)$ on either side of $\nu=1/4$\cite{Pan00}.

\begin{acknowledgments}
ACB was supported in part by the European Research Council (ERC) under the European Union Horizon 2020 Research and Innovation Programme, Grant Agreement No. 678862; by the Villum Foundation; and by the Center for Quantum Devices funded by the Danish National Research Foundation. JKJ was supported by the U. S. National Science Foundation Grant no. DMR-1401636. Some calculations were performed with Advanced CyberInfrastructure computational resources provided by The Institute for CyberScience at The Pennsylvania State University. Some of the numerical calculations were performed using the DiagHam package, for which we are grateful to its authors.
\end{acknowledgments}

\appendix
\section{2-body composite fermion pseudopotentials}
\label{app:two_body_CF_pps}
In Tables~\ref{tab:two_body_CFP_pps} and \ref{tab:two_body_CFH_pps} we show the 2-body pseudopotentials of CF particles and CF holes residing in different $\Lambda$Ls indexed by $n$. For $m>m_{\rm max}$ we set:
\begin{equation}
V^{\rm CF(2)}_{m}=\frac{1}{(2n+1)^{\frac{5}{2}}}
\frac{\Gamma(m+\frac{1}{2})}{2\Gamma(m+1)}~ \frac{e^{2}}{\epsilon\ell},~~~m>m_{\rm max}
\end{equation}
so that the pseudopotentials smoothly merge with the long-range Coulomb value.

\begin{table*}
\begin{center}
\begin{tabular}{|c|c|c|c|c|}
\hline
\multicolumn{1}{|c|}{$m$} & \multicolumn{1}{|c|}{$V^{{\rm CF(2)},n=0\downarrow}_{m}$ $(N=29)$}
& \multicolumn{1}{|c|}{$V^{{\rm CF(2)},n=1}_{m}$ $(N=32)$}
& \multicolumn{1}{|c|}{$V^{{\rm CF(2)},n=2}_{m}$ $(N=44)$} 
& \multicolumn{1}{|c|}{$V^{{\rm CF(2)},n=3}_{m}$ $(N=38)$} \\ \hline
1	&	0.0181(1)	&	0.0083(1)	&	-0.0056(4)	&	-0.0061(4)	\\	\hline
3	&	0.0140(1)	&	0.0216(1)	&	0.0034(4)	&	-0.0005(4)	\\	\hline
5	&	0.0111(2)	&	0.0067(1)	&	0.0075(4)	&	0.0039(4)	\\	\hline
7	&	0.0095(2)	&	0.0093(1)	&	-0.0008(5)	&	0.0067(4)	\\	\hline
9	&	0.0087(2)	&	0.0096(2)	&	0.0017(5)	&	-0.0008(6)	\\	\hline
11	&	0.0078(2)	&	0.0078(2)	&	0.0013(5)	&	0.0012(4)	\\	\hline
13	&	0.0074(2)	&	0.0070(2)	&	0.0015(4)	&	0.0006(5)	\\	\hline
15	&	0.0068(2)	&	0.0068(2)	&	0.0017(5)	&	0.0010(5)	\\	\hline
17	&	0.0067(2)	&	0.0068(2)	&	0.0010(4)	&	0.0009(0)	\\	\hline
19	&	0.0065(2)	&	0.0063(2)	&	0.0009(5)	&	0.0009(0)	\\	\hline
21	&	0.0064(2)	&	0.0061(2)	&	0.0016(5)	&	0.0008(0)	\\	\hline
23	&	0.0062(2)	&	0.0061(2)	&	0.0019(5)	&	0.0008(0)	\\	\hline
25	&	0.0064(2)	&	0.0060(2)	&	0.0018(0)	&	0.0008(0)	\\	\hline
27	&	0.0061(0)	&	0.0057(2)	&	0.0017(0)	&	0.0007(0)	\\	\hline
29	&	0.0059(0)	&	0.0057(2)	&	0.0017(0)	&	0.0007(0)	\\	\hline
31	&	0.0057(0)	&	0.0057(2)	&	0.0016(0)	&	0.0007(0)	\\	\hline
33	&	0.0056(0)	&	0.0056(0)	&	0.0016(0)	&	0.0007(0)	\\	\hline
35	&	0.0054(0)	&	0.0054(0)	&	0.0015(0)	&	0.0006(0)	\\	\hline
37	&	0.0053(0)	&	0.0053(0)	&	0.0015(0)	&	0.0006(0)	\\	\hline
39	&	0.0051(0)	&	0.0051(0)	&	0.0014(0)	&	0.0006(0)	\\	\hline
41	&	0.0050(0)	&	0.0050(0)	&	0.0014(0)	&	0.0006(0)	\\	\hline
43	&	0.0049(0)	&	0.0049(0)	&	0.0014(0)	&	0.0006(0)	\\	\hline
45	&	0.0048(0)	&	0.0048(0)	&	0.0013(0)	&	0.0006(0)	\\	\hline
47	&	0.0047(0)	&	0.0047(0)	&	0.0013(0)	&	0.0006(0)	\\	\hline
49	&	0.0046(0)	&	0.0046(0)	&	0.0013(0)	&	0.0005(0)	\\	\hline
\end{tabular}
\end{center}
\caption {The 2-body pseudopotentials of CF particles in the $n$th $\Lambda$L as a function of the relative angular momentum $m$. The energy are given in units of $e^{2}/(\epsilon\ell)$, calculated from a finite system of $N$ electrons in the spherical geometry. The value of $N$ is shown in each case in the top row. 
The number shown in the parenthesis is the error bar obtained from the Monte Carlo uncertainty. Some of these pseudopotentials were previously obtained by Lee \emph{et al.} \cite{Lee01,Lee02}} 
\label{tab:two_body_CFP_pps} 
\end{table*}

\begin{table*}
\begin{center}
\begin{tabular}{|c|c|c|c|c|c|}
\hline
\multicolumn{1}{|c|}{$m$} & \multicolumn{1}{|c|}{$V^{{\rm CF(2)},n=0\downarrow}_{m}$ $(N=50)$}
& \multicolumn{1}{|c|}{$V^{{\rm CF(2)},n=0}_{m}$ $(N=27)$}
& \multicolumn{1}{|c|}{$V^{{\rm CF(2)},n=1}_{m}$ $(N=60)$}
& \multicolumn{1}{|c|}{$V^{{\rm CF(2)},n=2}_{m}$ $(N=64)$} 
& \multicolumn{1}{|c|}{$V^{{\rm CF(2)},n=3}_{m}$ $(N=50)$} \\ \hline
1	&	0.0077(3)	&	0.0415(1)	&	-0.0021(6)	&	-0.0054(6)	&	-0.0038(5)	\\	\hline
3	&	0.0046(3)	&	0.0107(2)	&	0.0120(7)	&	0.0016(5)	&	-0.0002(5)	\\	\hline
5	&	0.0032(4)	&	0.0159(2)	&	0.0009(8)	&	0.0069(6)	&	0.0034(5)	\\	\hline
7	&	0.0029(4)	&	0.0097(2)	&	0.0025(7)	&	0.0002(6)	&	0.0074(5)	\\	\hline
9	&	0.0022(6)	&	0.0091(2)	&	0.0024(7)	&	0.0000(8)	&	0.0014(5)	\\	\hline
11	&	0.0022(4)	&	0.0087(2)	&	0.0033(8)	&	0.0013(8)	&	0.0010(5)	\\	\hline
13	&	0.0019(4)	&	0.0079(2)	&	0.0023(8)	&	0.0008(7)	&	0.0007(6)	\\	\hline
15	&	0.0020(3)	&	0.0073(2)	&	0.0011(8)	&	0.0004(8)	&	0.0005(6)	\\	\hline
17	&	0.0021(4)	&	0.0068(2)	&	0.0019(9)	&	0.0006(8)	&	0.0005(0)	\\	\hline
19	&	0.0018(4)	&	0.0067(2)	&	0.0018(9)	&	0.0006(8)	&	0.0005(0)	\\	\hline
21	&	0.0020(4)	&	0.0065(2)	&	0.0019(7)	&	0.0019(9)	&	0.0004(0)	\\	\hline
23	&	0.0015(5)	&	0.0064(2)	&	0.0016(9)	&	0.0008(7)	&	0.0004(0)	\\	\hline
25	&	0.0018(5)	&	0.0063(2)	&	0.0021(8)	&	0.0008(0)	&	0.0004(0)	\\	\hline
27	&	0.0017(0)	&	0.0061(2)	&	0.0013(9)	&	0.0007(0)	&	0.0004(0)	\\	\hline
29	&	0.0017(0)	&	0.0059(0)	&	0.0021(8)	&	0.0007(0)	&	0.0004(0)	\\	\hline
31	&	0.0016(0)	&	0.0057(0)	&	0.0016(9)	&	0.0007(0)	&	0.0004(0)	\\	\hline
33	&	0.0016(0)	&	0.0056(0)	&	0.0016(0)	&	0.0007(0)	&	0.0004(0)	\\	\hline
35	&	0.0015(0)	&	0.0054(0)	&	0.0015(0)	&	0.0006(0)	&	0.0003(0)	\\	\hline
37	&	0.0015(0)	&	0.0053(0)	&	0.0015(0)	&	0.0006(0)	&	0.0003(0)	\\	\hline
39	&	0.0014(0)	&	0.0051(0)	&	0.0014(0)	&	0.0006(0)	&	0.0003(0)	\\	\hline
41	&	0.0014(0)	&	0.0050(0)	&	0.0014(0)	&	0.0006(0)	&	0.0003(0)	\\	\hline
43	&	0.0014(0)	&	0.0049(0)	&	0.0014(0)	&	0.0006(0)	&	0.0003(0)	\\	\hline
45	&	0.0013(0)	&	0.0048(0)	&	0.0013(0)	&	0.0006(0)	&	0.0003(0)	\\	\hline
47	&	0.0013(0)	&	0.0047(0)	&	0.0013(0)	&	0.0006(0)	&	0.0003(0)	\\	\hline
49	&	0.0013(0)	&	0.0046(0)	&	0.0013(0)	&	0.0005(0)	&	0.0003(0)	\\	\hline
\end{tabular}
\end{center}
\caption {Same as Table~\ref{tab:two_body_CFP_pps} but for CF holes.} 
\label{tab:two_body_CFH_pps} 
\end{table*}

\bibliography{../../Latex-Revtex-etc./biblio_fqhe}
\bibliographystyle{apsrev}
\end{document}